\documentclass[prl,twocolumn]{revtex4-2}

\usepackage{graphicx}% Include figure files
\usepackage{dcolumn}% Align table columns on decimal point
\usepackage{bm}% bold math
\usepackage{mathrsfs}
\usepackage{subfigure}
\usepackage{natbib}
\usepackage{amsmath}
\usepackage{amssymb}
\usepackage{Jonasmacros}
\usepackage{mathptmx}
\usepackage{txfonts}
\usepackage{ifsym}

\usepackage[colorlinks,citecolor=magenta,linkcolor=red]{hyperref}
\usepackage[usenames,dvipsnames,svgnames]{xcolor}

\begin{document}

\title{Chiral induced Spin Polarized Electron Current: Origin of the Chiral Induced Spin Selectivity Effect}

\author{Jonas Fransson}
\email{Jonas.Fransson@physics.uu.se}
\affiliation{Department of Physics and Astronomy, Box 516, 75120, Uppsala University, Uppsala, Sweden}

%\significancestatement{
%The spin degree of freedom carried by electrons is in general coupled to the electron's spatial motion through spin-orbit coupling. As a consequence of this coupling, electrons moving in a chiral potential are subject to a spin-torque which tends to spin-polarize the electrons. The mechanism generating the spin-polarization is present in any chiral structure, even in the absence of external magnetic fields. In this paper it is shown that a steady-state spin-polarization can only arise when the electrons are undergoing spin-dependent dissipative processes. The spin-dependent dissipation allows the electrons to exchange angular momentum with its surroundings, either lowering or increasing its spin angular momentum. Furthermore, the mechanisms described are necessary and sufficient conditions for the chiral induced spin selectivity effect.
%}

%\keywords{chirality $|$ dissipation $|$ spin-polarization $|$ chiral induced spin selectivity}

\begin{abstract}
The discovery of the chiral induced spin selectivity effect has provided a novel tool to study how active physical and chemical mechanism may differ in chiral enantiomers, however, the origin of the effect itself is yet an open question. In this article, it is theoretically shown that two aspects have to be fulfilled for the chiral induced spin selectivity effect to arise. First, chirality is a necessary condition for breaking spin-degeneracy in molecular structures that do not comprise heavy elements. Second, dissipation is indispensable for the molecule to develop a non-vanishing spin-polarization. These theoretical conclusions are illustrated in terms of a few examples, showing the necessity of the two aspects to be coordinated for the emergence of the chiral induced spin selectivity effect.
\end{abstract}

\maketitle

Chirality induced spin selectivity is now a well-established phenomenon in physics, chemistry, and in biology \cite{ChemRev.124.1950}. Whenever electrons flow through a chiral material,  they will likely be spin-polarised in the direction they propagate \cite{Science.283.814}. Aside from its intrinsic interest, this phenomenon has implications both for devices and for our understanding of biological electrochemistry. In aerobes large electron currents, typically tens of amperes in the resting human, flow from metabolism to oxygen \cite{JChemEduc.60.289}. What makes spin important in this process of respiration is that the ultimate electron acceptor, dioxygen, is a ground state triplet. 

One of the reactions involved in the reduction of oxygen to water involves a two-electron transfer. It has been shown that this reaction is facilitated by spin-polarisation \cite{PNAS.2202650119}. At a ferromagnetic electrode, oxygen reduction proceeds faster in a magnetic field. Remarkably, general anesthetics, known to affect cellular respiration, markedly reduce spin-polarisation at a ferromagnetic electrode \cite{JPhysChemLett.14.1756}. This said, the extent and importance of spin-polarisation in living systems remains unknown. Direct two-terminal measurements of electron current in biology  are usually impossible. 

Hitherto, almost all measurements related to the chiral induced spin selectivity effect have thus far been made by injecting a spin-polarized current into the chiral material, e.g., see Refs \cite{NanoLett.11.4652,JPhysChemC.126.3257,JACS.144.7302}. In the case of photoemission, measurements using linearly polarized light were conducted \cite{Science.331.894}, demonstrating the spin-polarization of the electrons emitted from a surface of chiral molecules. By contrast, there has been little attempt to discern the magnetic properties of chiral structures in the absence of external magnetic boundary conditions. If that were possible, it may amount to a noninvasive method to assess for example whether the electron currents flowing in a living organism are spin-polarized. Recently, however, it was predicted that chiral molecules should acquire an inhomogeneous spin-distribution under a charge current flux \cite{JPhysChemLett.15.6370}

Theoretically, the chiral induced spin selectivity effect has been approached using the independent particle description \cite{PhysRevB.85.081404,PhysRevLett.108.218102,PNAS.111.11658,PhysRevB.93.075407,JPhysChemC.123.17043,PNAS.121.e2411427121}, which does not capture the chiral induced spin anisotropy that underlies the phenomenon. Density functional theoretical approaches have also been employed \cite{JChemTheoryComput.16.2914,JPhysChemLett.14.694}, however, without successfully accounting for these basic features of the effect. These approaches fail because the intrinsic absence of particle correlations that account for dissipation or losses in the system. It has subsequently been demonstrated that electron correlations are necessary to be included in order obtain a somewhat more physically correct description of the processes \cite{NanoLett.19.5253,JPhysChemLett.10.7126,PhysRevB.102.235416,JACS.143.14235,JPhysChemC.125.23364,JPhysChemC.127.6900,JChemPhys.158.174108,JPhysChemLett.14.340,NanoLett.24.12133}.

The prediction made in this article is that any charge current flowing through a chiral molecule must become spin-polarized. Behind the generation of the spin-polarized current is the breaking of time-reversal symmetry by dissipative processes and the breaking of spin-degeneracy by chirality coupled to spin-orbit coupling. In this sense one may describe chiral molecules as being spin-polarizers. The mechanism presented in this article plays an important role for the theoretical development of chiral induced spin selectivity effect.

Spin-polarization will emerge if two symmetries are broken, namely (i) time-reversal symmetry and (ii) spin-degeneracy. It is not necessary for a single mechanism to break the two symmetries simultaneously. There might be different sources for the two symmetry breaking agents. One should keep in mind that breaking time-reversal symmetry allows but does not imply breaking spin-degeneracy. Another important aspect is that the discussion of breaking time-reversal symmetry does not necessarily apply to the system as a whole, but only to parts that locally may correspond to lowering of the entropy as a local order is established. This loss of entropy is, however, compensated by an excess of entropy in another part of the system. Consequently, the description of local properties may display a broken symmetry state although the global symmetry is preserved.

An example is the screening that occurs around a localized magnetic moment in a metallic environment at low temperatures, due to the Kondo effect. While the local moment tends to stabilize at low temperatures, the surrounding environment of itinerant electrons align anti-ferromagnetically with the local moment such that the overall magnetic moment vanishes. In this sense, the local order which is established by the formation of the localized moment is prevented from violating the time-reversal invariance by the Kondo screening cloud.

The general structure of the electronic spin-orbit coupling can be effectively written as $\bfv\cdot\bfsigma$, where the vector $\bfv$ is defined by the electric field $\bfE$ and the momentum operator $\bfp$, and $\bfsigma$ is the vector of Pauli matrices. For a centrally symmetric (wave-) functions, the electric field $\bfE$ and, hence, the spin-orbit coupling vanishes. The wavefunction is required to be distorted from a centrally symmetric form to generate a spin-orbit coupling.

Consider a molecular structure defined by a distribution of coplanar nuclei. The effective spin-orbit coupling for such a structure provides a hybridization between the spins of the planar form $\bfv\cdot\bfsigma=v_-\sigma_++v_+\sigma_-$, where $v_\pm=v_x\pm iv_y$ and $\sigma_\pm=\sigma_x\pm i\sigma_y$. For a Kramers doublet, the planar spin-orbit coupling breaks the degeneracy of the state, however, into bonding and anti-bonding linear combinations of the spin-states such that the resulting gapped electron spectrum remains non spin-polarized. In principle this symmetry breaking mechanism does not have an influence on the spin-degeneracy, which remains intact.

By contrast, for a non-coplanar distribution of nuclei, the effective vector field may include a longitudinal component $v_z$. This leads to the introduction of a mass term $v_z\sigma_z$ in the matrix defining the spectrum. The significance of this mass term is to open up an energy gap between the spin-states, hence breaking the spin-degeneracy. It is important to point out that although this spin-degeneracy breaking mechanism exists in a molecule, it is an immanent property of the molecule. As such, it requires additional conditions to be fulfilled before an actual spin-polarization develops. Nevertheless, for chiral molecules, this mechanism has the effect to enable the emergence of a molecular spin-polarization under certain conditions that will be discussed next. 

The most important of those conditions is breaking time-reversal symmetry. Assume that the electronic structure of a molecular compound is captured by the single-electron Green's function $\mathbb{G}$. Assuming that the Green's function can be expressed in closed form as a Dyson equation, $\mathbb{G}=\mathbb{G}_0+\mathbb{G}_0\bfSigma\mathbb{G}$, where $\mathbb{G}_0$ and $\bfSigma$ represent the unperturbed Green's  function and self-energy, respectively. This is done in the paradigm in which the model Hamiltonian can be partitioned into non-interacting and interacting components. In such a representation, the density of electron states, ${\rm DOS}$, is generally defined according to
\begin{align}
{\rm DOS}=&
	i~{\rm tr}\Bigl(
		\mathbb{G}^>-\mathbb{G}^<
	\Bigr)
	,
\label{eq-DOSlg}
\end{align}
where the lesser and greater forms $\mathbb{G}^{</>}$ are proportional to the density of occupied and unoccupied electrons states, respectively. However, the density of electron states can equally well be formulated in terms of the retarded and advanced forms $\mathbb{G}^{r/a}$, thanks to the fundamental equality $\mathbb{G}^>-\mathbb{G}^<=\mathbb{G}^r-\mathbb{G}^a$. Then, the Dyson equation allows to reformulate Eq. \eqref{eq-DOSlg} as
\begin{align}
{\rm DOS}=&
	i~{\rm tr}
	\mathbb{G}^r
	\Bigl(
		\bfSigma^r-\bfSigma^a
	\Bigr)
	\mathbb{G}^a
	,
\end{align}
since $\mathbb{G}^r_0-\mathbb{G}^a_0=0$.

Keeping in mind that the Green's function is the matrix $\mathbb{G}=\{\bfG_{mn}\}_{mn}$ over the $N$ electronic states. The element $\bfG_{mn}$ is a $2\times2$-matrix representing the electron propagation between states $m$ and $n$. In terms of the these $2\times2$-matrices, the expression can be further expanded as
\begin{align}
{\rm DOS}=&
	i~{\rm sp}
	\sum_{kmn}
	\bfG^r_{km}
	\Bigl(
		\bfSigma^r_{mn}-\bfSigma^a_{mn}
	\Bigr)
	\bfG^a_{nk}
	,
\label{eq-DOSGSG}
\end{align}
where ${\rm sp}$ denotes the trace over spin 1/2 space.

Relating this expression to time-reversal symmetry is done by studying its properties under the time-reversal operator $\mathbb{T}=i\sigma^rK$, where $K$ denotes complex conjugation. Under the trace, the Pauli matrix $\sigma^y$ commutes with the summand in Eq. \eqref{eq-DOSGSG}, however, the complex conjugation in general does  not, since $K(\bfSigma^r_{mn}-\bfSigma^a_{mn})=-(\bfSigma^r_{nm}-\bfSigma^a_{nm})K$. This expression is non-zero whenever the self-energy comprises an imaginary part. Since the imaginary part of the self-energy corresponds to dissipative processes in the system, the physical implication of this statement is that dissipation is a manifestation of broken time-reversal symmetry.

Note here that the discussion about the broken time-reversal symmetry has to do with the subsystem to which special attention is paid. The full system, which is a closed entity, is always time-reversal symmetric. However, the full properties of the subsystem are determined not only by its intrinsic configuration but also by the environment with which it interacts and exchanges, e.g., energy, momentum, and angular momentum. In these interactions, losses which are lost to the environment are inevitable.

The combination of mechanisms that break time-reversal symmetry and spin-degeneracy leads to an induced spin-polarization in chiral molecules. This conclusion can be demonstrated through a model of a vibrating molecule attached to metallic leads and embedded in a thermal reservoir, represented by the Hamiltonian $\Hamil=\Hamil_L+\Hamil_R+\Hamil_\text{therm}+\Hamil_\text{mol}+\Hamil_T$. Here, $\Hamil_\chi=\sum_{\bfk\in\chi}\dote{\bfk}\psi^\dagger_\bfk\psi_\bfk$ defines free electrons at the energy $\dote{\bfk}$ and wave vector $\bfk$ in the lead $\chi=L,R$ in terms of the spinor $\psi_\bfk=(\psi_{\bfk\up}\ \psi_{\bfk\down})^t$. The thermal reservoir is defined by the harmonic oscillators $\Hamil_\text{therm}=\sum_\bfq\omega_\bfq b^\dagger_\bfq b_\bfq$, where $b_\bfq$ and $b^\dagger_\bfq$ annihilates and creates a thermal phonon at the energy $\omega_\bfq$ and wave vector $\bfq$. The vibrating molecule is described by \cite{PhysRevB.102.235416}
\begin{subequations}
\begin{align}
\Hamil_\text{mol}=&
	\sum_{mn}\psi^\dagger_m\Bigl(E+H_0+H_1\sum_\nu\omega_\nu(a_\nu+a^\dagger_\nu)\Bigr)_{mn}\psi_n
\nonumber\\&
+
	\Hamil_\text{vib}
	,
\\
\Bigl(H_l\Bigr)_{mn}=&
	\sum_{s=\pm1}
	\Bigl(
		-t_l\delta_{nm+s}
		+
		i\lambda_l\bfv_m^{(s)}\cdot\bfsigma\delta_{nm+2s}
	\Bigr)
    ,
\nonumber\\&
	l=0,1,
\end{align}
\end{subequations}
and $(E)_{mn}=\delta_{mn}\dote{m}$. The electrons in the molecule are distributed among $\mathbb{M}$ sites at the on-site energies $\dote{m}$ and elastically hybridized with the nearest and next-nearest neighbors with rates $t_0$ and $\lambda_0$, as well as vibrationally assisted nearest and next-nearest hybridization with rates $t_1$ and $\lambda_1$. The next-nearest hybridization comprises the spin-orbit coupling $\bfv_m^{(s)}\cdot\bfsigma$, where $\bfv_m^{(s)}=\hat{\bfd}_{m,s}\times\hat{\bfd}_{m,s+1}$, $\bfd_{m,s}=\bfr_m-\bfr_{m+s}$, and $\hat{\bfd}_{m,s}=\bfd_{m,s}/|\bfd_{m,s}|$. The vector $\bfv_m^{(s)}$ defines the curvature between the sites $\bfr_m$, $\bfr_{m+s}$, and $\bfr_{m+2s}$. The nuclear vibrations are represented by harmonic oscillations in $\Hamil_\text{vib}=\sum_\nu\omega_\nu a^\dagger_\nu a_\nu$, with energies $\omega_\nu$ created and annihilated by the phonon operator $a_\nu$ and $a^\dagger_\nu$, respectively.

Finally, the molecule is coupled to the leads and thermal reservoir through
\begin{align}
\Hamil_T=&
	\sum_{\bfk\in L}
		\Bigl(
			v_{\bfk 1}\psi^\dagger_\bfk\psi_1
			+
			H.c.
		\Bigr)
	+
	\sum_{\bfk\in R}
		\Bigl(
			v_{\bfk \mathbb{M}}\psi^\dagger_\bfk\psi_\mathbb{M}
			+
			H.c.
		\Bigr)
\nonumber\\&
	+
	\sum_{\bfq\nu}\Phi_{\bfq\nu}(b_\bfq+b^\dagger_{\bar\bfq})(a_\nu+a^\dagger_\nu)
	,
\end{align}
where $v_{\bfk m}$, $m=1,\mathbb{M}$, is the tunneling rate for electrons between the molecule and the leads, whereas $\Phi_{\bfq\nu}$ denotes the coupling rate between the nuclear vibrations and thermal phonons, and $\bar\bfq=-\bfq$.

\begin{figure}[t]
\begin{center}
\includegraphics[width=\columnwidth]{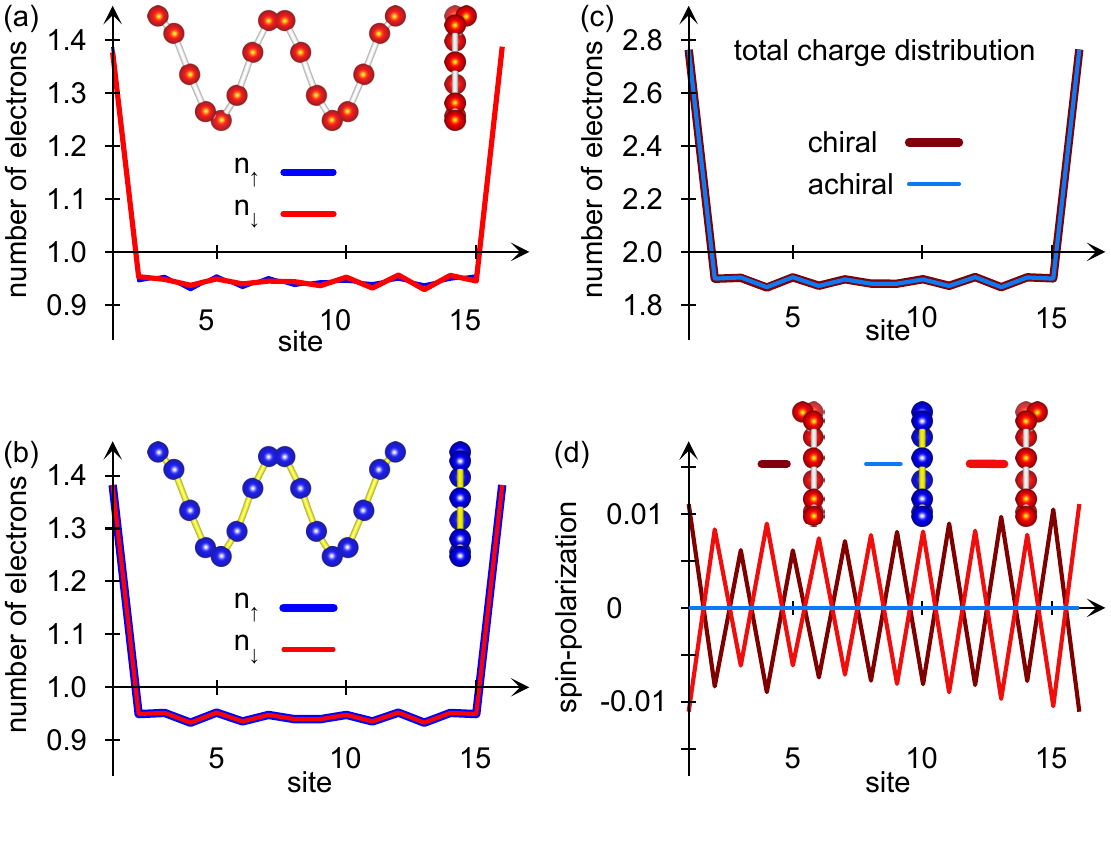}
\end{center}
\caption{Calculated equilibrium spin-resolved charge distributions for (a) a chiral and (b) an achiral molecular chain comprising $\mathbb{M}=16$ sites. The total charge distributions $n_\sigma$, $\sigma=\up,\down$, for both cases are shown in panel (c), whereas the corresponding spin-polarizations $n_\up-n_\down$ are plotted in panel (d), where the spin-polarization for the opposite enantiomer is also shown. Here, $\dote{m}=-50$, $t_1=1/10$, $\lambda_0=1/25$, $\lambda_1=1/250$, $\Gamma_\chi=1/2$, and $\omega_\nu=1/100,000$, in units of $t_0=0.1$ eV, at the temperature $T=300$ K.}
\label{fig-ChiralAchiralCD}
\end{figure}

From this model, the electronic structure can be calculated in terms of the single-electron Green's function $\bfG_{mn}$, which describes the electron propagation between sites $m$ and $n$. To the second order in the vibrationally assisted hybridization $\lambda_l$, this Green's function is given by \cite{PhysRevB.102.235416}
\begin{subequations}
\begin{align}
\mathbb{G}(z)=&
	\biggl(
		z-E-H_0
		-
		H_1^2\Sigma_\text{ph}(z)
	\biggr)^{-1}
	,
\\
\Sigma_\text{ph}(z)=&
	\sum_\nu
		\biggl(
			\frac{n_B(\omega_\nu)+1-f(\dote{0})}{z-\dote{0}-\omega_\nu+i/\tau_\text{ph}}
			+
			\frac{n_B(\omega_\nu)+f(\dote{0})}{z-\dote{0}+\omega_\nu+i/\tau_\text{ph}}
		\biggr)
	,
\end{align}
\end{subequations}
where $\dote{0}=\sum_m\dote{m}/\mathbb{M}$. Here, the life-time $\tau_\text{ph}$ arises from the coupling between the nuclear vibrations and thermal reservoir and explicitly accounts for  the losses associated with inelastic processes. In this formulation, also the degrees of freedom associated with the leads are integrated out and captured by the addition of the level broadening $\Gamma_\chi/2$ to the energies $E_{1/\mathbb{M}}=\dote{1/\mathbb{M}}-i\Gamma_{L/R}/2$, where $\Gamma_\chi=2\pi\sum_{\bfk\in\chi}|v_{\bfk m}|^2\delta(\omega-\dote{\bfk m})$.

\begin{figure}[t]
\begin{center}
\includegraphics[width=\columnwidth]{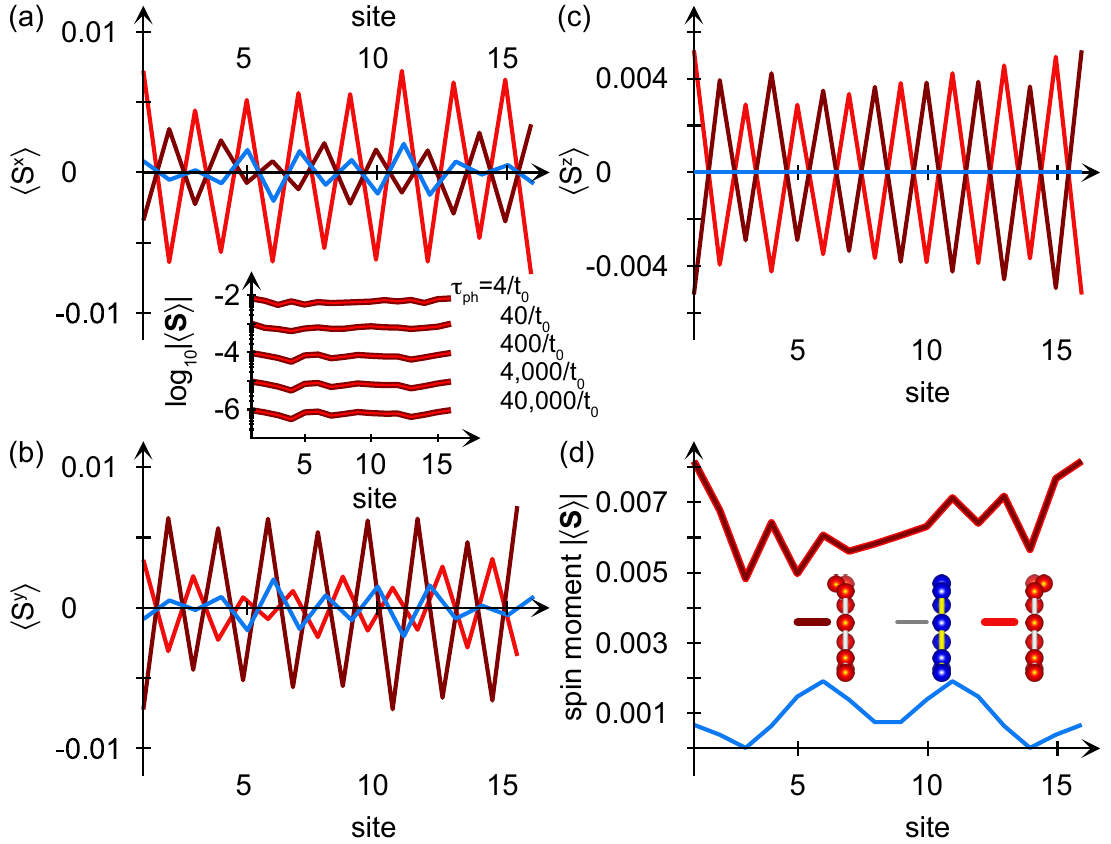}
\end{center}
\caption{Calculated equilibrium spin-distributions projected in (a) $\av{S^x}$, (b) $\av{S^y}$, (c) $\av{S^z}$, and (d) $|\av{\bfS}|$ for an (blue) achiral and (red, dark red) two-enantiomers of chiral molecular chains comprising $\mathbb{M}=16$ sites. The inset between panels (a) and (b) shows the induced total spin moment $|\av{\bfS}|$ for increasing life-time $\tau_\text{ph}$. Other parameters are as in Fig. \ref{fig-ChiralAchiralCD}.}
\label{fig-ChiralAchiralSD}
\end{figure}

Numerical results by applying the model to chiral and achiral molecules are shown in Figs. \ref{fig-ChiralAchiralCD} and \ref{fig-ChiralAchiralSD} for structures schematically illustrated in the insets of Fig. \ref{fig-ChiralAchiralCD} (a) and (b). The achiral structure is a planar zig-zag chain, Fig. \ref{fig-ChiralAchiralCD} (b), whereas the chiral deviates from planarity in one site that is non-coplanar with the zig-zag structure. As predicted from the theoretical discussion above, the internal spin-degeneracy is broken, see Fig. \ref{fig-ChiralAchiralCD} (a), (d), which show the spin resolved charge distribution $n_\sigma$, $\sigma=\up,\down$ and spin-polarization $n_\up-n_\down$. The plots in fig. \ref{fig-ChiralAchiralCD} (d), moreover, illustrates the opposite spin-polarizations arising in the two enantiomers. For the achiral structure, on the other hand, the charge distributions for the two spin-projections are equal, Fig. \ref{fig-ChiralAchiralCD} (b), which implies a vanishing spin-polarization, Fig. \ref{fig-ChiralAchiralCD} (d).

However, despite the absence of a spin-polarization in the achiral structure, there is nonetheless a non-trivial spin-texture, which can be seen in Fig. \ref{fig-ChiralAchiralSD} (a)--(c), in which the expectation value of the spin-projections $\av{S^\alpha}$, $\alpha=x,y,z$ are plotted, as well as in the total spin moment $|\av{\bfS}|$, Fig. \ref{fig-ChiralAchiralSD} (d). While the longitudinal spin-projection $\av{S_z}=(n_\up-n_\down)/2$ vanishes, the spin-orbit coupling induces a transverse spin-texture $\av{S^{x,y}}\neq0$. The origin of this texture can be found in the curvature vector $\bfv_m^{(s)}$ which is necessarily perpendicular to the longitudinal spin orientation $\sigma^z$. However, for non-collinear achiral structures, the curvature vector $\bfv_m^{(s)}$ is coplanar with $\sigma^{x,y}$.

By contrast, in chiral structures the curvature vector $\bfv_m^{(s)}$ also comprises a component in the longitudinal direction which breaks the chiral symmetry of the structure. Therefore, since the orbital quality is coupled to the spin through the spin-orbit coupling, the chiral symmetry breaking also breaks the spin-degeneracy, which can be seen in Fig. \ref{fig-ChiralAchiralSD}. It is, moreover, important to notice that helicity is not necessary in this context but the more general character chirality is what makes the difference. This is also known from experimental observations of the chiral induced spin selectivity effect on, e.g., cysteine and similar compounds \cite{PNAS.2202650119}.

The inset in Fig. \ref{fig-ChiralAchiralSD}, in which the total spin moment $|\av{\bfS}|$ is plotted for different vibrational life time $\tau_\text{ph}$, illustrates the importance of dissipative processes for the induced spin-polarization in this context. The magnitude of the spin-moment decreases nearly an order of magnitude for each order of magnitude increase in the vibrational life-time. This clearly indicates that (i) the chirality broken spin-degeneracy does not lead to a spin-polarization in a closed molecule, despite it being coupled to external leads, and (ii) the thermal environment has to be included in a full description of the chiral induced spin selectivity effect.

That there is an induced spin-polarization in chiral structures in equilibrium is a prerequisite for the chiral induced spin-selectivity to arise. In absence of such intrinsic spin anisotropy, the chiral enantiomers cannot respond anisotropically to an injected spin-polarized current.

\begin{figure}[t]
\begin{center}
\includegraphics[width=\columnwidth]{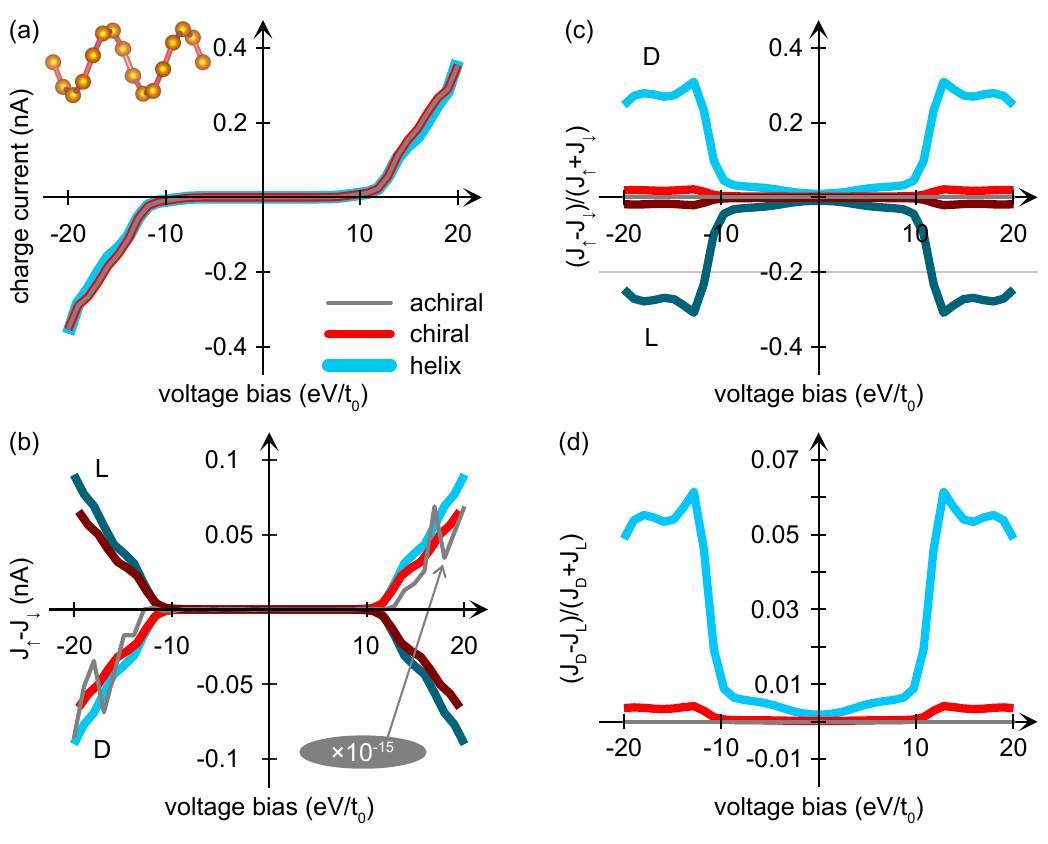}
\end{center}
\caption{Transport properties for the (gray) achiral and (red) chiral structures depicted in Fig. \ref{fig-ChiralAchiralCD}, as well as (blue) helical structures. The latter comprise $\mathbb{M}=16$ sites equidistantly distributed with respect to the azimuthal angle $\varphi\in[0,4\pi]$, as depicted in the inset of panel (a). The panels show the (a) charge current, (b) current spin-polarization, (c) current spin-polarization normalized by the charge current, and (d) the enantiomeric asymmetry when injecting a 20 \% spin-polarized current from one of the leads. In panels (b) and (c), the chiral enantiomers are assigned either $L$ or $D$. In panel (c), the faint line at $-0.2$ on the vertical axis represents zero spin-polarization for the conditions in  panel (d). Other parameters are as in Fig. \ref{fig-ChiralAchiralCD}.}
\label{fig-SpinPolarizedCurrent}
\end{figure}

The transport properties of the achiral and chiral structures depicted in Fig. \ref{fig-ChiralAchiralCD} are plotted in Fig. \ref{fig-SpinPolarizedCurrent}, in which the (a) charge current and (b) current spin-polarization are shown as function of the voltage bias across the junction. The current gap for voltage biases between $\pm10$ reflects the energetic distance between the equilibrium chemical potential and the nearest (occupied) molecular orbitals. Despite that the charge currents are equal in the (gray) achiral and (red) chiral structures, the latter constitute a non-vanishing current spin-polarization, as expected, see Fig. \ref{fig-SpinPolarizedCurrent} (b). These plots also demonstrate that the current spin-polarization is enantiospecific, that is mirror symmetric with respect to enantiomer L or D, and is of the order of several percents of the charge current. This is illustrated by the plots of the ratio $(J_\up-J_\down)/(J_\up+J_\down)$ in Fig. \ref{fig-SpinPolarizedCurrent} (c).

The chiral structures depicted in Fig. \ref{fig-ChiralAchiralCD} are minimally chiral, in the sense that only one site is non-coplanar with all other sites. While this quality is sufficient to generate the spin-dependent properties observed for chiral structures, a comparison is made with a helical geometry in which the same number of site (16) are equidistantly distributed with respect to the azimuthal angle $\varphi\in[0,4\pi]$, see inset of Fig. \ref{fig-SpinPolarizedCurrent} (a). The charge current (blue) of this structure is nearly the same as for the other two geometries, however, the current spin-polarization that arises is significantly stronger, see Fig. \ref{fig-SpinPolarizedCurrent} (b), (c). Especially, as the voltage bias reaches the molecular orbitals and the current rises, it is accompanied by a strong spin-polarization from which it can be concluded that the molecular electronic structure needs to actively participate in the conduction through the molecule for a strong spin-polarization to arise. This is, however, obvious since the gap between orbitals, which essentially is vacuum, cannot generate any spin-signatures.

The emergence of a current spin-polarization explains the results obtained in photo-emission experiments \cite{Science.331.894}, where spin-polarized photo-excited currents were detected using Mott scattering. Detecting a spin-polarized intensity can only be the result of a spin-imbalance of the electrons which are ejected from the chiral molecules. In this experiment, electrons in the Au substrate are photoexcited and forced to be transported in the unoccupied orbitals in double stranded DNA molecules. By applying the arguments used here for empty orbitals for a molecule attached to a single metals, leads to the qualitatively same result as shows in Fig. \ref{fig-SpinPolarizedCurrent} (b), (c).

\begin{figure}[t]
\begin{center}
\includegraphics[width=\columnwidth]{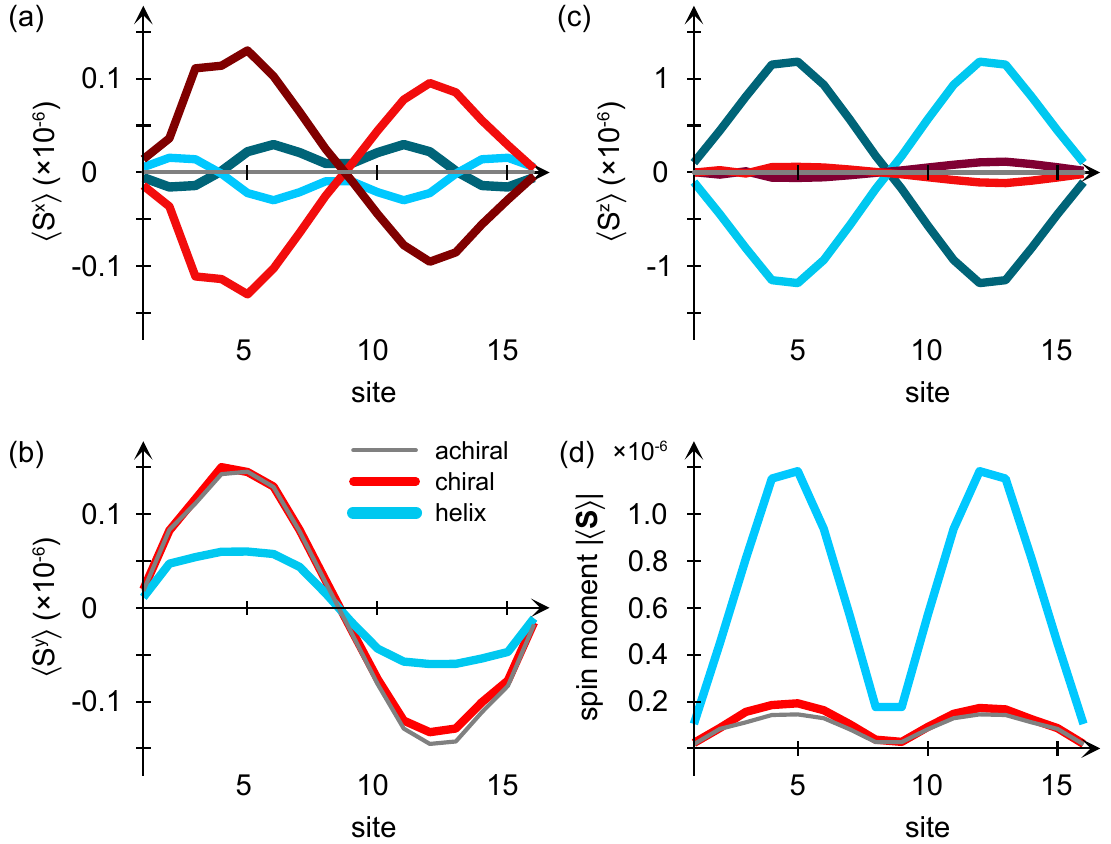}
\end{center}
\caption{Calculated equilibrium spin-distributions projected in (a) $\av{S^x}$, (b) $\av{S^y}$, (c) $\av{S^z}$, and (d) $|\av{\bfS}|$ for an (blue) achiral and (red, dark red) two-enantiomers of chiral molecular chains comprising $\mathbb{M}=16$ sites. The plots represent calculations with on-site Coulomb repulsion. Here, $\dote{m}=-2$, $U=1/5$, $\lambda_0=1/100$, and $\Gamma_\chi=1$, in units of $t_0=0.1$ eV, at the temperature $T=300$ K.}
\label{fig-ChiralAchiralSDCoulomb}
\end{figure}

In the transport configuration, on the other hand, the chiral induced spin selectivity effect is quantified by, for instance, comparing the charge currents $J_D$ and $J_L$ measured for the $D$ and $L$ enantiomers under the injection of an externally spin-polarized current, respectively \cite{NanoLett.11.4652}. The intrinsically generated chiral induced spin-polarization explains the phenomenon of the chiral induced spin selectivity effect in this set-up, since there is an immanent spin anisotropy that can be enhanced by an externally injected spin-polarization, provided that the spin-polarizations are concurrent. If instead the spin-polarizations are competing, the molecular spin anisotropy is suppressed and eventually coerced to align with the external spin-polarization when this is sufficiently strong. The externally induced shift of the immanent spin-polarization is indicated by the faint line at $-0.2$ on the vertical axis in Fig. \ref{fig-SpinPolarizedCurrent} (c). Effectively, the current spin-polarization is shifted upwards compared to zero spin-polarization by injecting a 20 \% spin-polarized current from one of the leads. In the calculation this is achieved by replacing, e.g., $\Gamma_L\rightarrow\bfGamma_L=\Gamma_L(\sigma^0+p_L\sigma^z)/2$, where $p_L=0.2$ defines the spin-polarization of the injected electrons.

Then, comparing the resulting charge currents for the two enantiomers through the ratio $(J_D-J_L)/(J_D+J_L)$, the  results for the chiral (red) and helical (blue) structures are plotted in Fig. \ref{fig-SpinPolarizedCurrent} (d). Both the weakly chiral and the helical structures display this enantiomeric asymmetry which has become a hallmark for the chiral induced spin selectivity effect in the transport configuration. The weakly chiral structure also has a weaker enantiomeric asymmetry than the helical, which can be understood as a large phase accumulation in the helical structure due to its repeated curvature property.

It is important to remark that the chiral induced spin-polarization and the chiral induced spin selectivity effect does not depend on the origin of the interactions and dissipation in the system. Replacing the vibrational and phonon degrees of freedom by electrostatic Coulomb repulsion at each site in the molecular chain results in the analogous results, see Ref. \cite{JPhysChemLett.10.7126} for details, summarized for a realistic example in Fig. \ref{fig-ChiralAchiralSDCoulomb}.

Similarly as in the example with vibrations, the achiral structure (gray) acquires a non-trivial spin-texture, Fig. \ref{fig-ChiralAchiralSDCoulomb} (b) and total spin moment Fig. \ref{fig-ChiralAchiralSDCoulomb} (d). Nevertheless, there is no longitudinal spin component, Fig. \ref{fig-ChiralAchiralSDCoulomb} (c). The chiral structures (red and blue), on the other hand, do acquire non-vanishing longitudinal spin components, Fig. \ref{fig-ChiralAchiralSDCoulomb} (c), (d), albeit small under the present conditions. The weak spin-polarization is related to that the Coulomb repulsion in the present example is defined as a local on-site interaction only, whereas the vibrational effects in the previous example is non-local by construction. Hence, an addition of non-local Coulomb interactions would enhance the induced spin-polarization.

In conclusion it has been demonstrated that dissipation in chiral structures leads to the formation of a non-trivial and inhomogeneous spin-density. Chirality is a manifestation of a structural broken symmetry which by spin-orbit coupling breaks the spin-degeneracy. However, the formation of a spin-density requires a broken time-reversal symmetry, and since the formation of a spin-asymmetry, or, spin-density is also identical to the establishment of an order, the entropy of the electronic structure has to be lowered. One implication of this statement is that any structure which does not possess a mechanism for spin-dependent losses cannot maintain a spin-density as a steady state. A second implication is that the origin of the spin-dependent dissipative processes is of subordinate importance. Finally, one consequence of spin-dependent dissipation in chiral structures is the chiral induced spin selectivity effect.

\acknowledgements
The author is grateful to A. Sisman and R. Naaman for fruitful and inspiring discussions. A special thanks to Luca Turin for generously spending hours in \emph{rambling discussions}. Support from Stiftelsen Olle Engkvist Byggm\"astare is acknowledged.

%\showacknow{} % Display the acknowledgments section

%\bibsplit[2]
%Use \bibsplit to split the references from the body of the text. Value "[2]" represents the number of reference in the left column (Note: Please avoid single column figures & tables on this page.)

% Bibliography
\bibliography{CISSref}

\end{document}